\documentclass[conference]{IEEEtran}

\usepackage{cite} 
\usepackage{graphicx} 
\usepackage{amsmath} 
\usepackage{amssymb} 
\usepackage{algorithmic} 
\usepackage{textcomp} 
\usepackage{url} 
\usepackage{subcaption}
\usepackage{booktabs}
\usepackage{tikz}
\usetikzlibrary{arrows.meta,angles,quotes,calc}

\begin{document}

\title{See and Beam: Leveraging LiDAR Sensing and Specular Surfaces for Indoor mmWave Connectivity}

\author{
    Raj Sai Sohel Bandari, Amod Ashtekar,  Omar Ibrahim,  and Mohammed E. Eltayeb \\
    Department of Electrical and Electronic Engineering, \\
    California State University, Sacramento, Sacramento, USA \\
    Email: \{rbandari, amodashtekar, omaribrahim2,  mohammed.eltayeb\}@csus.edu
}

\maketitle

\begin{abstract}
Millimeter‐wave (mmWave) communication enables multi-gigabit-per-second data rates but is highly susceptible to path loss and blockage, especially indoors. Many indoor settings, however, include naturally occurring \emph{specular surfaces} such as glass, glossy metal panels,  and signage,  that reflect both light and mmWave signals. Exploiting this dual reflectivity, we propose \textit{See and Beam}, a low-cost framework that combines LiDAR sensing with passive specular reflectors to enhance mmWave connectivity under non-line-of-sight (NLoS) conditions.  In this paper, as a proof of concept, we deploy three types of reflectors,  glossy,  smooth,  and matte (non-specular),  to evaluate joint LiDAR/mmWave reflection in an indoor scenario. We demonstrate that using LiDAR–mmWave co-reflective surfaces enables a co-located LiDAR sensor to map the NLoS environment, localize NLoS users, and identify viable communication reflection points. Experimental results at 60 GHz show that LiDAR-guided beam steering with co-reflective surfaces improves the minimum received signal strength by over 20 dB in deep NLoS regions. Moreover, LiDAR-derived angle-of-departure steering achieves performance comparable to exhaustive NLoS beam search.  This  low cost, and scalable  framework serves as an effective alternative to configurable reflecting surfaces and enables robust mmWave connectivity in future 6G and beyond networks.
\end{abstract}


\section{Introduction}
The growing demand for immersive video, extended‐reality, and massive-IoT applications is pushing wireless systems toward \emph{multi-gigabit-per-second} data rates. The millimeter-wave (mmWave) spectrum offers the requisite bandwidth and has already been embraced by 5G NR while featuring prominently in 6G roadmaps\,\cite{alsabah2021survey,Rappaport2019B5G,3gpp20235g}.
Yet the short wavelength that delivers such capacity also leads to {severe path loss and weak diffraction}, making mmWave links highly vulnerable to blockage, particularly indoors, where walls, furniture, and users frequently obstruct the line-of-sight (LoS) path\,\cite{topal2024}.

To mitigate these limitations, recent work investigated various physical-layer and environmental solutions, such as beamforming, hybrid arrays, reconfigurable intelligent surfaces (RIS), and passive reflectors~\cite{qian2022millimirror, khawaja2020coverage, PR3}. Among these, passive reflectors offer a promising approach due to their low cost, lack of energy requirements, and compatibility with existing infrastructure. Metallic reflectors, in particular, provide specular reflection and can redirect mmWave signals around obstacles to enable extended coverage without the need for active elements.  Yet,  prior work often assumes optimal reflector orientation and seldom considers user mobility\,\cite{khawaja2020coverage,qian2022millimirror,PR3}.

\begin{figure}[t]
    \centering

    \begin{subfigure}[b]{0.15\textwidth}
        \includegraphics[width=\textwidth]{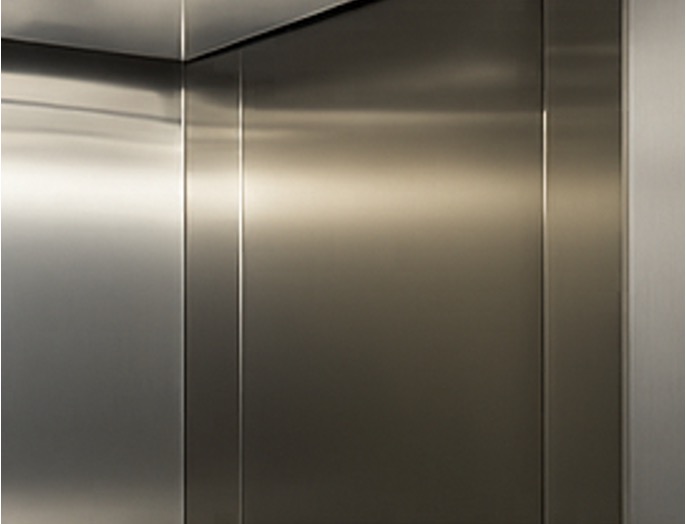}
        \caption{Polished stainless steel}
    \end{subfigure}
    \hfill
    \begin{subfigure}[b]{0.15\textwidth}
        \includegraphics[width=\textwidth]{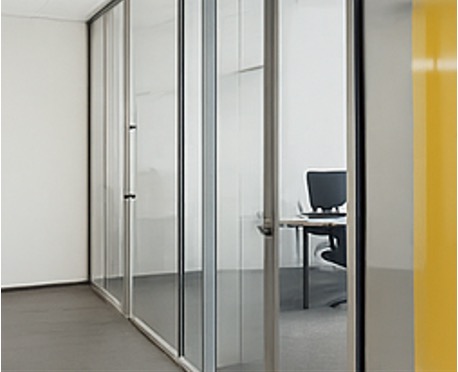}
        \caption{Glass partition}
    \end{subfigure}
    \hfill
    \begin{subfigure}[b]{0.15\textwidth}
        \includegraphics[width=\textwidth]{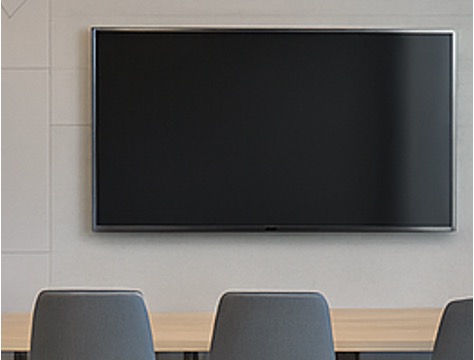}
        \caption{TV/monitor screen}
    \end{subfigure}

    \vskip\baselineskip

    \begin{subfigure}[b]{0.15\textwidth}
        \includegraphics[width=\textwidth]{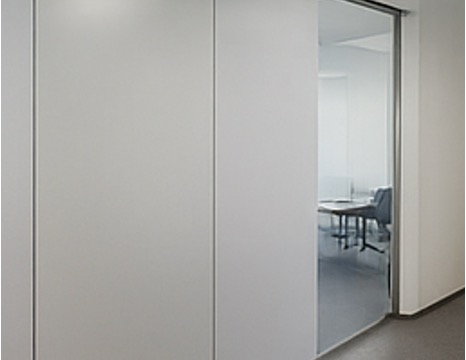}
        \caption{Aluminum composite panel}
    \end{subfigure}
    \hfill
    \begin{subfigure}[b]{0.15\textwidth}
        \includegraphics[width=\textwidth]{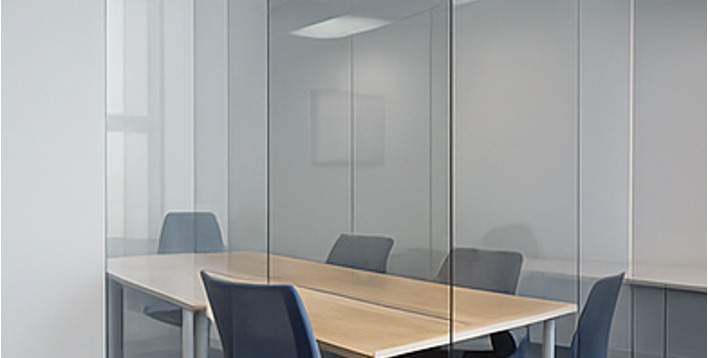}
        \caption{Acrylic/polycarbonate sheet}
    \end{subfigure}
    \hfill
    \begin{subfigure}[b]{0.15\textwidth}
        \includegraphics[width=\textwidth]{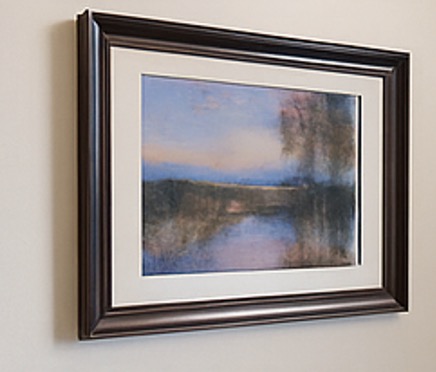}
        \caption{Framed artwork (glass)}
    \end{subfigure}

    \vskip\baselineskip

    \begin{subfigure}[b]{0.15\textwidth}
        \includegraphics[width=\textwidth]{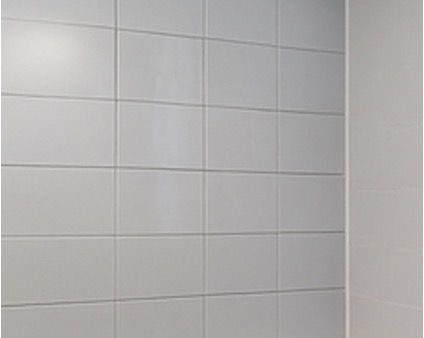}
        \caption{Glossy ceramic tile}
    \end{subfigure}
    \hfill
    \begin{subfigure}[b]{0.15\textwidth}
        \includegraphics[width=\textwidth]{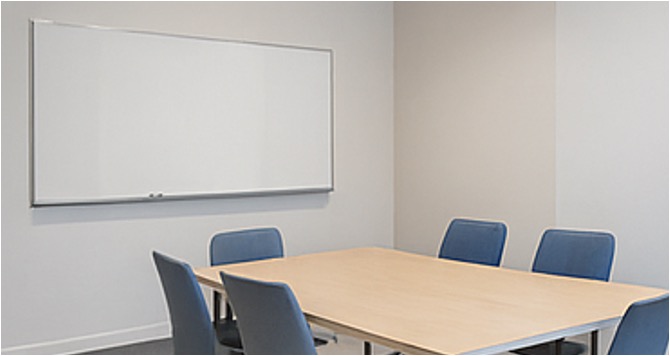}
        \caption{Glossy whiteboard panel}
    \end{subfigure}
    \hfill
    \begin{subfigure}[b]{0.15\textwidth}
        \includegraphics[width=\textwidth]{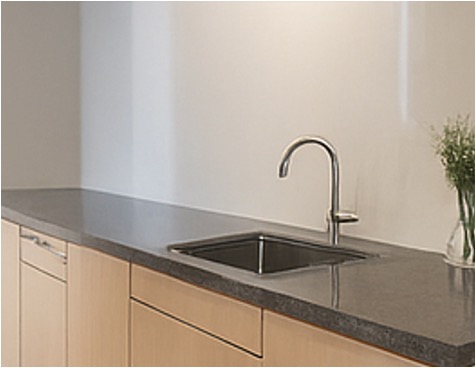}
        \caption{Polished stone countertop}
    \end{subfigure}
    
       \vskip\baselineskip

    \begin{subfigure}[b]{0.15\textwidth}
        \includegraphics[width=\textwidth]{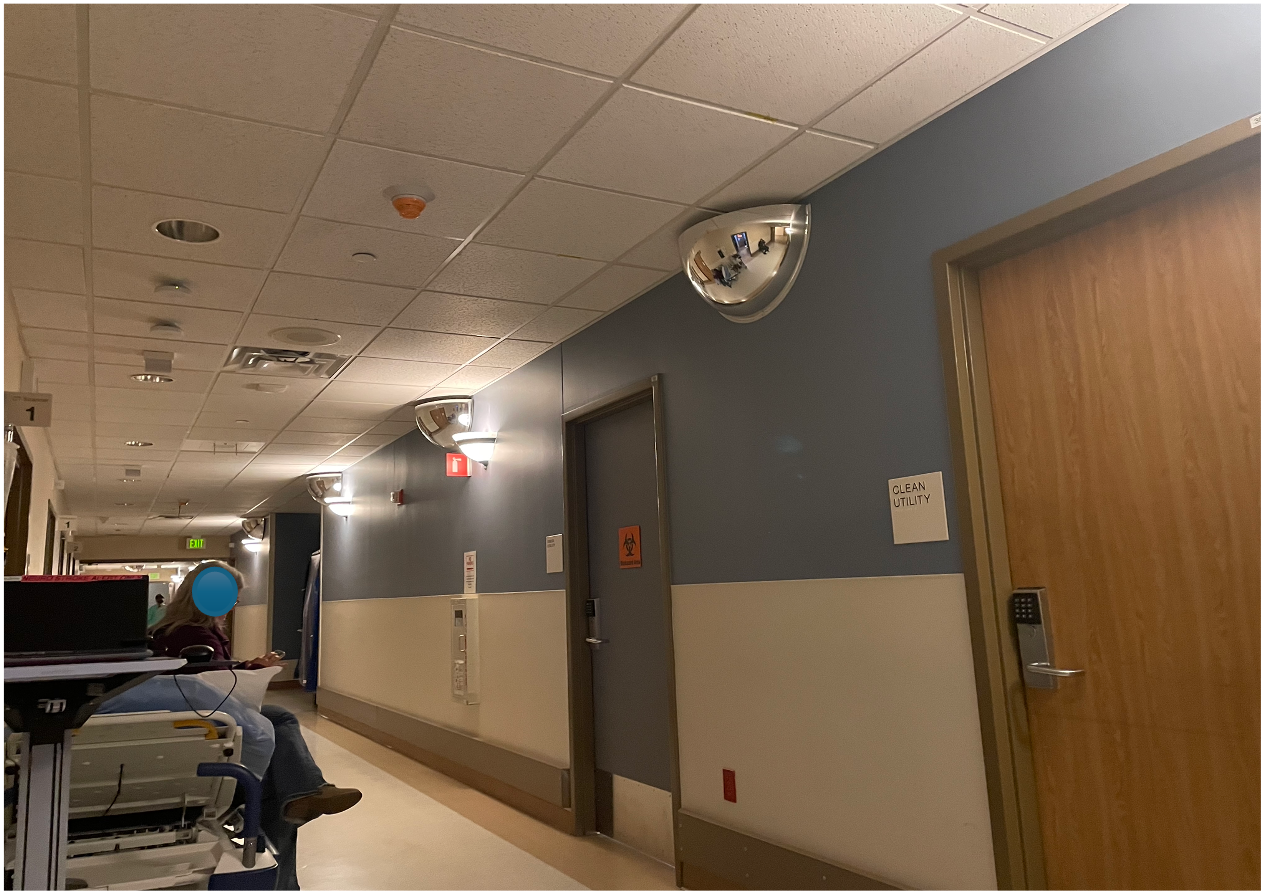}
        \caption{Blindspot mirror}
    \end{subfigure}
    \hfill
    \begin{subfigure}[b]{0.15\textwidth}
        \includegraphics[width=\textwidth]{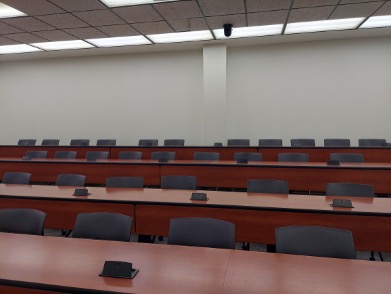}
        \caption{Polished wood}
    \end{subfigure}
    \hfill
    \begin{subfigure}[b]{0.15\textwidth}
        \includegraphics[width=\textwidth]{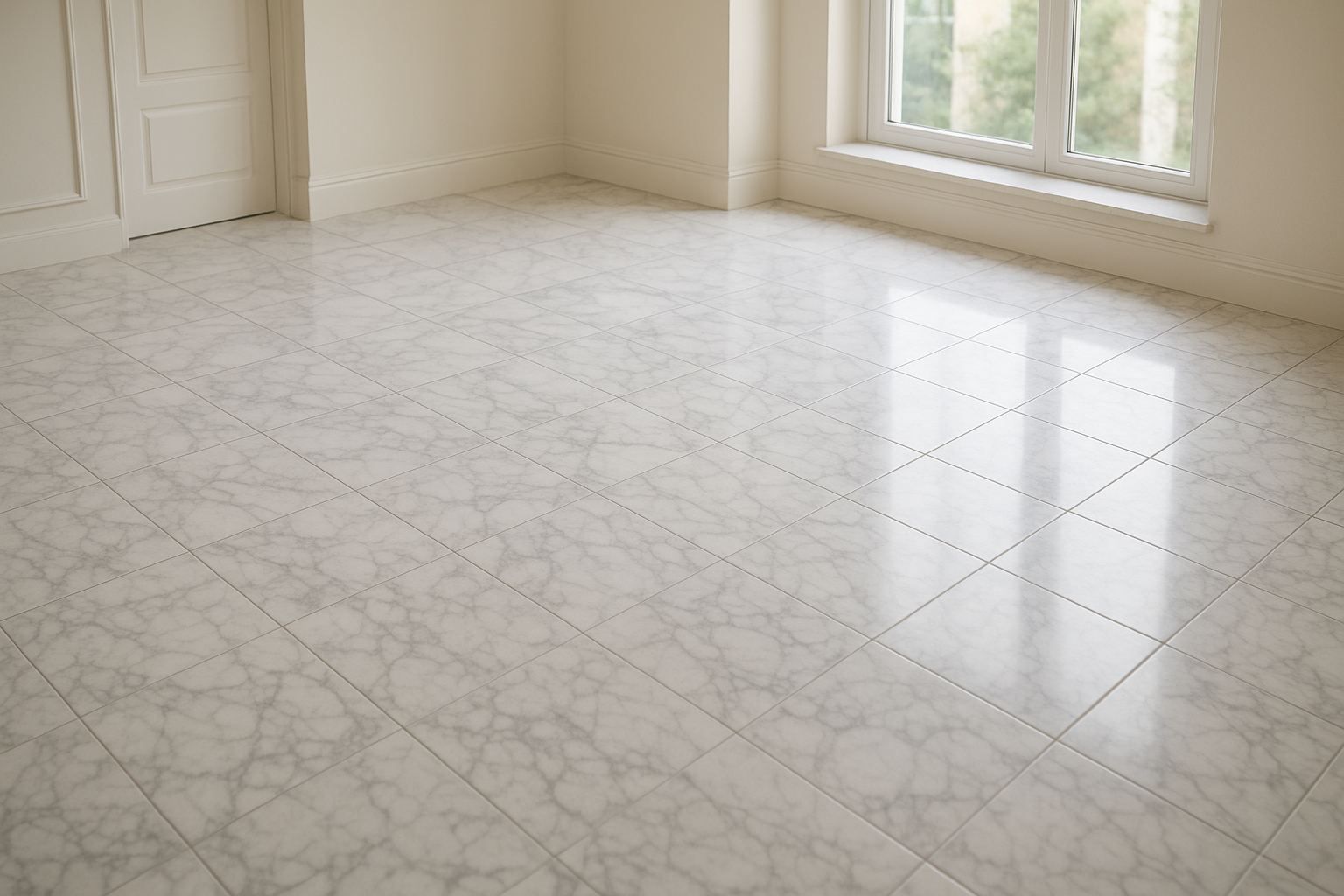}
        \caption{Marble flooring}
    \end{subfigure}
    
    \caption{Indoor surface types with specular or semi-specular reflection properties relevant to mmWave and LiDAR sensing. These materials can serve as passive mmWave reflectors and optical relays for joint sensing and NLoS communication.}
    \label{fig:specular_surfaces}
\end{figure}

Indoor spaces are naturally \emph{abound with specular surfaces} such as glass walls, polished metal panels, glossy signage,  and TV screens  that can reflect both light and mmWave energy \cite{spacebeam2021,ju2019scattering,Langen1994,Rappaport2015} (see Fig. \ref{fig:specular_surfaces}).   Concurrently, low-cost solid-state \emph{LiDAR} sensors now deliver centimeter-level 3-D maps at video frame rates.  
These two facts point to a new direction: sense the environment with LiDAR at optical wavelengths, then steer mmWave beams along the discovered specular routes, \emph{without} any active RF infrastructure in the environment.

In this paper, we present \textit{See and Beam}, a practical framework that integrates a commodity LiDAR sensor with passive specular/semi-specular reflectors to establish robust non-line-of-sight (NLoS) mmWave links.  In contrast to reconfigurable intelligent surface solutions,  passive reflectors can be found naturally in indoor environments and require no power, cabling, or embedded RF electronics. In environments lacking natural reflectors, a dual-reflector unit can be temporarily mounted on walls or structural pillars to extend both mmWave and LiDAR coverage. This enables mmWave transmitters to opportunistically beamform toward NLoS users based on their real-time location, while also optimizing \textit{reflector efficiency} by leveraging the most effective section(s) of the reflector surface.   The contributions of this work are threefold: (i) we formulate joint LiDAR–mmWave coordination as a two-stage sensing and beam-steering problem using low-cost  co-reflective surfaces; (ii) we demonstrate that mmWave specular paths can be directly inferred from LiDAR point clouds when using co-reflective surfaces; and (iii) we show that optimizing reflection points i.e., selecting optimal visibility regions on the reflector using LiDAR,  improves mmWave coverage in NLoS areas. 

\section{A Primer on Reflection and Scattering Losses}

Electromagnetic (EM) wave interaction with indoor surfaces is governed by the principles of reflection and scattering, both of which depend strongly on material permittivity, surface roughness, and wave incidence angle. Many architectural surfaces, such as glazed ceramic tiles, polished glass, aluminium cladding, or stainless-steel fixtures (see Fig.~\ref{fig:specular_surfaces}), serve as effective passive reflectors for both \emph{mmWave} and \emph{LiDAR}-based systems~\cite{Rappaport2019B5G,spacebeam2021}. Whether a given surface reflects more efficiently in one band than the other is primarily determined by (i) its complex permittivity at the operational frequency,  (ii) its local surface roughness relative to wavelength,  and (iii) the angle and polarization of incidence.

\subsection{Fresnel Reflection at Smooth Interfaces}
When an electromagnetic (EM) wave encounters a smooth boundary between two different media, the proportion of energy that is reflected versus transmitted at the interface is described by the \textit{Fresnel equations}. These equations provide the \textit{reflection coefficients} for the electric field components that are \textit{perpendicular} ($\perp$) and \textit{parallel} ($\parallel$) to the plane of incidence. For a planar interface between two non-magnetic, homogeneous media, the reflection coefficients are given by~\cite{ju2019scattering, Balanis2012, Langen1994, landron1993}:
\begin{equation}
\Gamma_\perp = \frac{\eta_2 \cos\theta_i - \eta_1 \cos\theta_t}{\eta_2 \cos\theta_i + \eta_1 \cos\theta_t}, \quad
\Gamma_\parallel = \frac{-\eta_1 \cos\theta_i + \eta_2 \cos\theta_t}{\eta_1 \cos\theta_i + \eta_2 \cos\theta_t}.
\label{eq:fresnel_general}
\end{equation}

\noindent
The variables $\eta_1$ and $\eta_2$ in (\ref{eq:fresnel_general}) denote the intrinsic impedances of the incident and transmission media, respectively, which depend on the material permittivity and permeability. The angle $\theta_i$ is the angle of incidence, measured from the normal to the interface, and $\theta_t$ is the angle of transmission (or refraction) into the second medium, determined by Snell's law. These coefficients quantify the ratio of the reflected electric field amplitude to the incident electric field amplitude for each polarization component.

Assuming both media are non-magnetic, such that the relative permeability is approximately unity (\( \mu_r \approx 1 \)),  the general Fresnel equations yields simplified forms that depend only on the relative permittivities of the two media such that $\Gamma_\perp = \frac{\cos\theta_i - \sqrt{\frac{\varepsilon_{r2}}{ \varepsilon_{r1}} } \cos\theta_t}{ \cos\theta_i +  \sqrt{\frac{\varepsilon_{r2}}{ \varepsilon_{r1}} } \cos\theta_t}, 
\Gamma_\parallel = \frac{- \cos\theta_i +  \sqrt{\frac{\varepsilon_{r1}}{ \varepsilon_{r2}} } \cos\theta_t}{ \cos\theta_i +  \sqrt{\frac{\varepsilon_{r1}}{ \varepsilon_{r2}} } \cos\theta_t}.$

\noindent
Here, \( \varepsilon_{r1} \) and \( \varepsilon_{r2} \) denote the relative permittivities of the incident and transmission media, respectively. These expressions are particularly useful in high-frequency applications such as mmWave and optical domains, where permittivity dominates the electromagnetic behavior of materials.

\begin{table*}[t]
\caption{Examples of Indoor Surfaces with Reflection Properties and Dielectric Characteristics.  The loss tangent $\tan(\delta)$ quantifies dielectric energy loss.}
\label{tab:specular_surfaces}
\centering
\begin{tabular}{|p{3.5cm}|p{2.8cm}|p{6.2cm}|}
\hline
\textbf{Surface or Object} & \textbf{Location} & \textbf{Reflection Characteristics} \\
\hline
Polished stainless steel panels (conductive) & Elevators, lobbies, kitchenettes & Metal surface; very high reflectivity at both mmWave and LiDAR due to strong impedance mismatch \cite{Balanis2012}.  Surface smoothness leads to highly specular returns. \\
\hline
Glass partitions / doors (\( \varepsilon_r = 5.29,\, \tan\delta = 0.048 \)) & Offices, meeting rooms & Dielectric with moderate permittivity and loss; supports specular reflection at oblique angles. Reflectivity increases with incidence angle. \\
\hline
Gloss-painted drywall (\( \varepsilon_r \approx 2.8,\, \tan\delta \approx 0.016 \)) & Hallways, workspaces & Glossy coating gives smooth air–surface transition for LiDAR reflectivity; weak-to-moderate reflectivity at mmWave. \\
\hline
TV/monitor screens (glass + polymer) & Offices, desks & Multi-layer dielectric stack with glass cover; high reflectivity and smoothness cause strong specular returns in both mmWave and optical bands. \\
\hline
Aluminum composite panels (ACP) & Feature walls, decorative cladding & Metallic surface layer ensures excellent mmWave reflectivity. LiDAR reflection depends on surface polish and oxide layer; typically moderately specular. \\
\hline
Acrylic glass / polycarbonate (\( \varepsilon_r = 2.53,\, \tan\delta = 0.0119 \)) & Desk shields, signage & Low-loss dielectrics with smooth surface finish yield moderate Fresnel reflectivity and partial specular behavior, especially under oblique angles. \\
\hline
Ceramic tiles (\( \varepsilon_r = 6.30,\, \tan\delta = 0.0568 \)) & Bathrooms, corridors & High permittivity and polish provide strong mmWave reflection; LiDAR reflection depends on glaze finish and surface roughness. \\
\hline
Framed artwork (glass cover) & Offices, corridors & Glass surface behaves as a specular reflector for both LiDAR and mmWave, especially at shallow angles. Paper or canvas behind contributes diffuse scattering. \\
\hline
Whiteboards or glossy panels  & Meeting rooms & Smooth coated surfaces  yield strong LiDAR return and moderate mmWave reflection depending on surface type (typically melamine, glass or painted steel).  Gloss level and base material affect overall reflectivity. \\
\hline
Polished stone countertops (\( \varepsilon_r = 6.81,\, \tan\delta = 0.0401 \)) & Kitchens, lobbies & Dense, polished dielectric; reflects mmWave effectively at oblique angles and offers semi-specular LiDAR returns due to surface smoothness. \\
\hline
Dry wood surfaces (\( \varepsilon_r = 1.5\!-\!4,\, \tan\delta = 0.01 \)) & Furniture, flooring & Low permittivity and moderate absorption; weak mmWave reflectivity, though polished or coated wood may return modest LiDAR signals. \\
\hline
Plasterboard (\( \varepsilon_r = 2.81,\, \tan\delta = 0.0164 \)) & Interior walls & Lossy dielectric; generally poor LiDAR and mmWave reflectivity unless painted or coated. \\
\hline
\end{tabular}
\end{table*}

\subsection{Effect of Surface Roughness}

In real-world environments, surfaces are rarely perfectly smooth. When surface irregularities approach a significant fraction of the wavelength, the reflected energy is scattered into non-specular directions, reducing coherent return. The boundary between smooth and rough is described by the Rayleigh criterion $h_c = \frac{\lambda}{8 \cos\theta_i} $
where \( h_c \) is the critical height and \( \lambda \) is the wavelength \cite{ju2019scattering}.  If the root-mean-square surface height \( h_{\text{rms}} > h_c \), the surface is considered rough at that wavelength.

For relatively large grazing angles,  the specular reflection loss due to roughness can be quantified by a scattering loss factor \( \rho_s \), expressed as $\rho_s = \exp\left[-8 \left(\frac{\pi h_{\text{rms}} \cos\theta_i}{\lambda}\right)^2\right]$ ~\cite{ju2019scattering, Langen1994}. 
The effective Fresnel reflection coefficients under rough surface conditions are therefore scaled as
\begin{equation}
\Gamma_{\perp,\text{rough}} = \rho_s \Gamma_\perp, \quad
\Gamma_{\parallel,\text{rough}} = \rho_s \Gamma_\parallel.
\label{eq:rough_fresnel}
\end{equation}
This model captures how materials with high intrinsic reflectivity (e.g., metal, tile) may yield poor LiDAR return if the surface microtexture causes optical scattering.

\subsection{Material-Specific Examples}

\begin{figure}[t]
    \centering
    \includegraphics[width=0.85\linewidth]{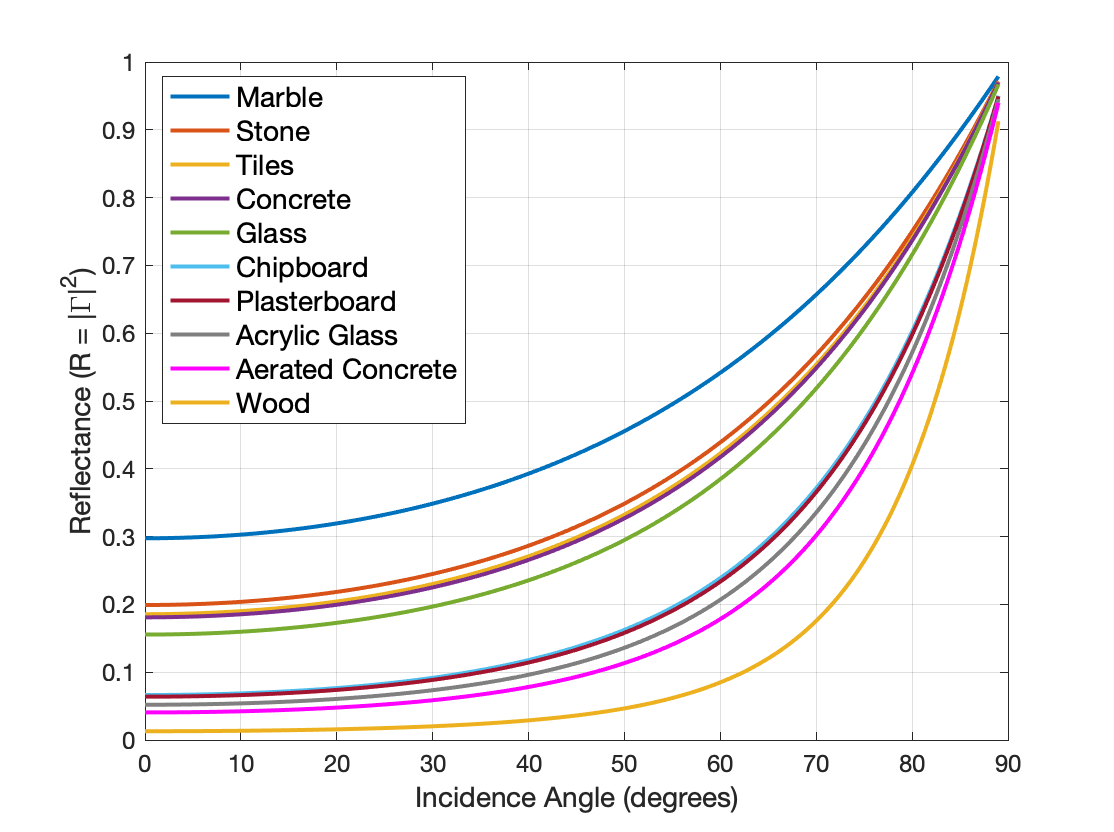}
    \caption{
    Simulated reflectance versus incidence angle at 60~GHz for selected indoor materials using their complex dielectric constants obtained from \cite{Correia1994}, and \cite{Balanis2012}.}
    \label{fig:reflectance_mmwave_sorted}
\end{figure}

Table~\ref{tab:specular_surfaces} summarizes a variety of commonly encountered indoor surfaces, highlighting their estimated or measured dielectric properties along with observed reflection characteristics for both mmWave and LiDAR signals.   These examples are based on (i) dielectric parameters from \cite{Correia1994,Balanis2012} and (ii) Fresnel-theory simulations, as shown in Fig.~\ref{fig:reflectance_mmwave_sorted}. The results in Fig.~\ref{fig:reflectance_mmwave_sorted} illustrate the s-polarized (parallel) reflectance profiles at 60~GHz for selected materials, assuming smooth, planar interfaces. The trends highlight how material permittivity governs mmWave reflection strength, providing a basis for selecting suitable reflector surfaces in practical deployments.
 While LiDAR reflectance at 905~nm can be modeled using the same theoretical framework, it is significantly more sensitive to surface roughness and microscopic scattering effects.  As illustrated in Table~\ref{tab:specular_surfaces}, materials such as ceramic tiles, polished stone, and glass partitions exhibit high reflectivity and strongly specular behavior, particularly at oblique angles. In contrast, surfaces like chipboard and unpainted drywall contribute minimal coherent return unless modified with a smooth or reflective coating.  Together, these measurements and simulations provide an understanding of how common indoor surfaces interact with both mmWave and LiDAR signals. 

\section{Power Model and Reflection Strategy}

\subsection{Power Model with Radar Cross Section}
Under far-field and specular reflection conditions, the received power $P_{\text{Rx}}$ at the NLoS receiver due to reflection from a passive reflector is given by the bistatic radar equation~\cite{Rappaport2002,radarcrossection} as $P_{\text{Rx}} = \frac{P_{\text{Tx}} G_{\text{Tx}} G_{\text{Rx}} \lambda^2 \sigma(\beta)}{(4\pi)^3 d_1^2 d_2^2},$
where $P_{\text{Tx}}$ is the transmit power, $G_{\text{Tx}}$ and $G_{\text{Rx}}$ are the antenna gains at the transmitter and receiver, respectively,  $d_1$ and $d_2$ are the distances from the transmitter to the reflector and from the reflector to the receiver, respectively.  The bistatic radar cross section (RCS)  $\sigma(\beta) \approx  \frac{4\pi A^2}{\lambda^2} \cos\left(\frac{\beta}{2}\right),$ where $A$ is the effective reflecting area and $\beta$ is the bistatic angle subtended at the reflector between the transmitter and receiver.

To incorporate the effect of the reflector's material properties,  we scale the RCS by the Fresnel power reflection coefficient $R_p(\theta_i, \varepsilon_r) = |\Gamma_{\perp,\text{rough}} |^2$ for parallel polarization to obtain
\begin{equation}
\sigma(\beta) \approx R_p(\theta_i, \varepsilon_r) \cdot \frac{4\pi A^2}{\lambda^2} \cos\left(\frac{\beta}{2}\right).
\end{equation}

This formulation shows that the received power is sensitive not only to the orientation of the reflector but also to its electromagnetic properties. The maximum received power is achieved when $\beta \to 0$, i.e., when the transmitter and receiver are symmetrically aligned about the surface normal, and the surface reflectivity is high (e.g., metallic or low-loss dielectric).  

\subsection{Reflection Control Strategy}
To optimize the received power, the proposed system uses LiDAR to minimize the bistatic angle $\beta$ by beamforming or changing the angle of incidence at the reflector such that the angles of incidence and reflection satisfy the condition $\theta_i\approx \theta_t \approx \beta/2$. This adaptive strategy significantly improves the NLoS link quality in complex or obstructed environments and can be further used to identify suitable user dependent visibility regions in a large reflector surface.   By leveraging this spatial information, the base station can dynamically select optimal reflection zones to support passive beam redirection for multiple users simultaneously.

\section{Experimental and Measurement Setup}
\begin{figure}[t]
    \centering

    \begin{subfigure}[b]{0.15\textwidth}
        \includegraphics[width=\textwidth]{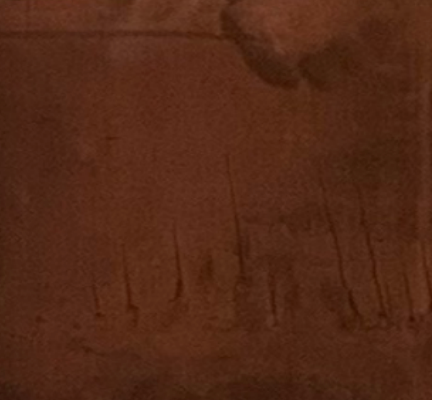}
        \caption{Copper surface}
    \end{subfigure}
    \hfill
    \begin{subfigure}[b]{0.13\textwidth}
        \includegraphics[width=\textwidth]{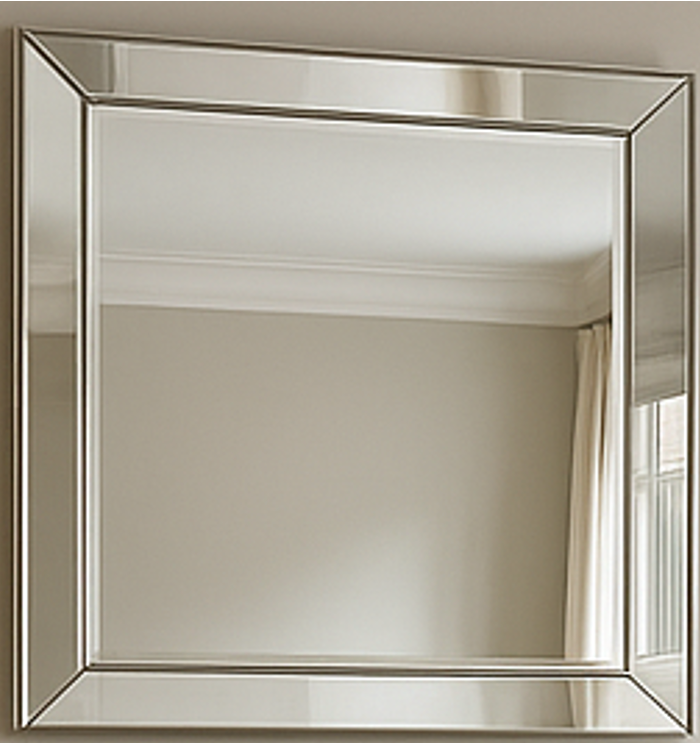}
        \caption{Mirror}
    \end{subfigure}
    \hfill
    \begin{subfigure}[b]{0.16\textwidth}
        \includegraphics[width=\textwidth]{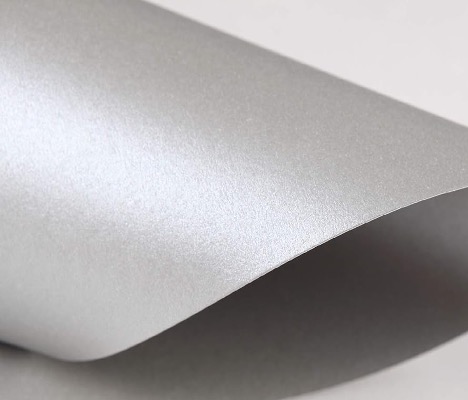}    
        \caption{Glossy silver sheet}
         \label{fig:UTsurfaces1}
    \end{subfigure}
    \caption{Flat reflector surfaces of size 0.3 × 0.9 $m^2$ mounted on a foam board
at an azimuth angle of $45^o$ with respect to the transmitter.}
    \label{fig:UTsurfaces}
\end{figure}

To validate the proposed LiDAR-guided passive reflection framework, we evaluate three surface types,  non-specular copper,  a mirror,  and a glossy silver sheet,  for their joint reflectivity under LiDAR and mmWave sensing, as shown in Fig.~\ref{fig:UTsurfaces}. The copper surface emulates materials such as aluminum composite panels commonly found indoors; it is highly reflective for mmWave signals but poorly reflective for LiDAR due to its surface roughness. The silver-coated mirror surface, is highly reflective for both modalities. The glossy silver sheet exhibits high mmWave reflectivity and moderate LiDAR reflectivity.  These surfaces span a representative range of reflectivity properties encountered in practical indoor environments.

\subsection{Measurement Environment}
Experiments were conducted in an L-shaped indoor corridor specifically chosen to eliminate direct line-of-sight and suppress multipath between the transmitter and receiver as shown in Fig.~\ref{fig:expsetup}.  The corridor is 2.5 meters wide, with the TX positioned 3.8 meters from the reflector and mounted at a height of 1.5 meters. A foam board placed at a 45-degree angle at the corner served as a mounting structure for the reflectors, which were flat rectangular panels with dimensions $0.3 \times 0.9~\text{m}^2$. The RX scanned the received signal over a uniform grid consisting of 102 points to capture spatial variations in signal strength across the NLoS region.

\subsection{Hardware Setup}
The measurement setup comprised two 60\,GHz mmWave transceiver kits (Sivers EVK02001 \cite{SiversIMA}), each equipped with an 8-element phased antenna array.  A Universal Software Radio Peripheral (USRP),  LiDAR sensor and synchronized laptops were used for control and data capture. The transmitter beam was electronically steered over angles ranging from $-15^\circ$ to $+15^\circ$ to illuminate different regions of the reflector, while the receiver beam remained fixed at $0^\circ$, aimed directly at the reflector to maximize received power. 

\begin{figure}[t]
    \centering
    \includegraphics[width=0.35\textwidth]{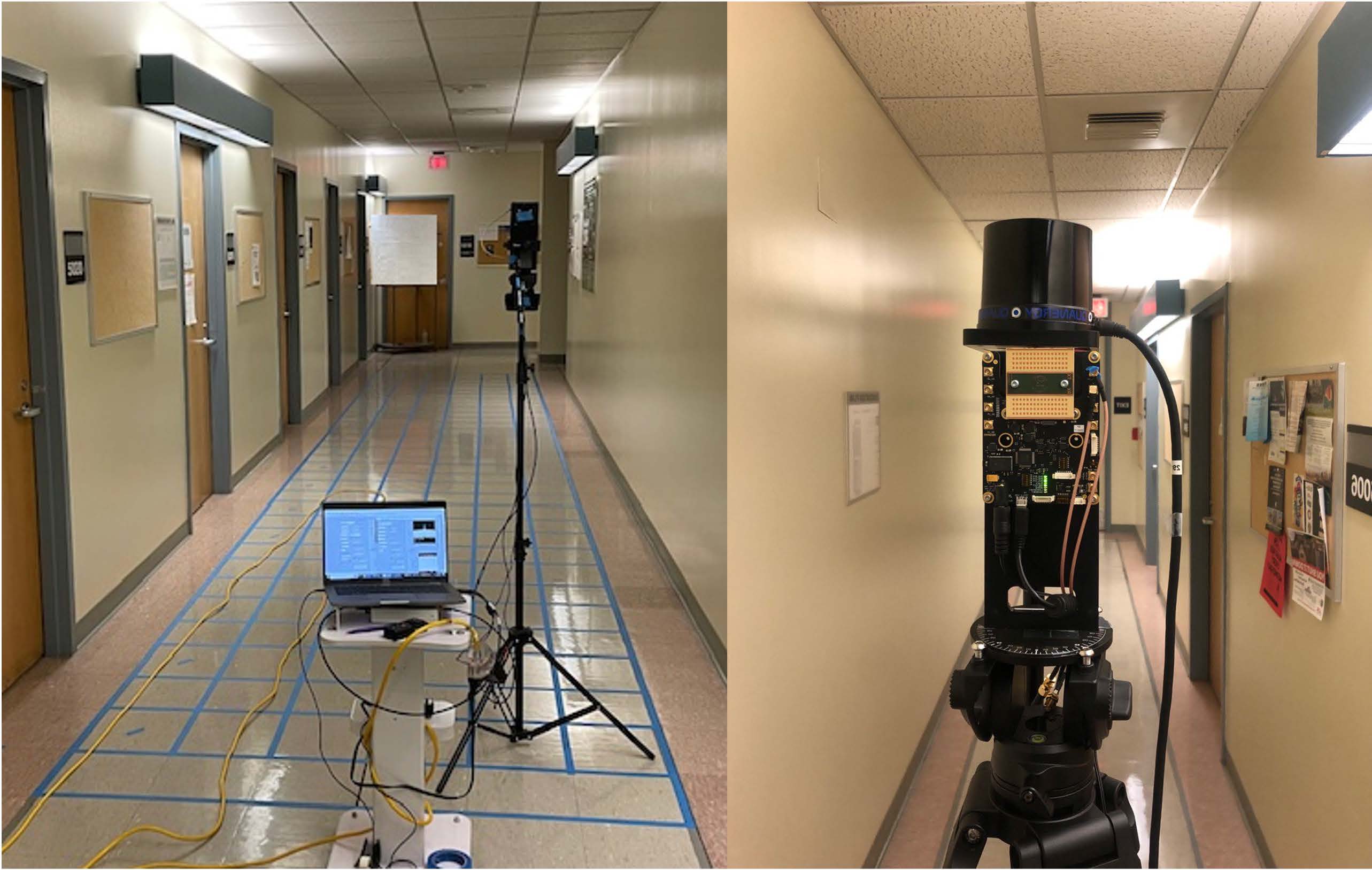}
    \caption{Left:  L-shaped corridor featuring 102 grid points for
RSS measurements. Right: image of the
co-located mm-wave transmitter and LiDAR at the end of the corridor.}
    \label{fig:expsetup}
\end{figure}

\begin{figure}[t]
    \centering
    \includegraphics[width=0.35\textwidth]{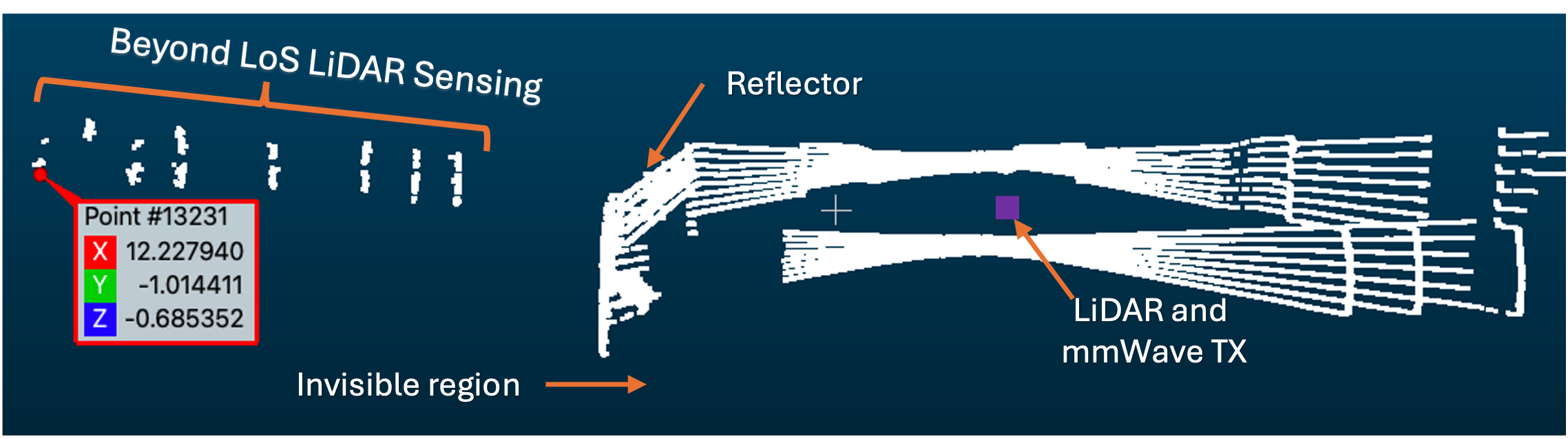}
    \caption{Beyond LoS LiDAR tracking of a user (1.8 m tall) in an L-shaped corridor using
a glossy silver sheet reflector (Fig. \ref{fig:UTsurfaces1}) of size $0.3 \times 0.9 m^2$.  } 
    \label{fig:lidar}
\end{figure}

LiDAR data were acquired using a commercial solid-state LiDAR sensor (Quanergy M8) and processed off-line in CloudCompare. For each receiver position, the LiDAR point cloud was used to estimate the angles of departure, $\theta_i$ and $\theta_t$, as well as the corresponding reflection point location on the reflector surface. These estimates were obtained by transforming the reflection point coordinates into the global reference frame, computing the incident and reflected vectors, and applying vector geometry to determine the angles. The estimated LiDAR-based angles of departure were then used to guide mmWave beam steering toward the reflector. The overall environment and hardware deployment are shown in Fig.~\ref{fig:expsetup}, and a representative LiDAR point cloud is presented in Fig.~\ref{fig:lidar}. LiDAR and mmWave measurements were conducted separately and fused offline to determine the optimal beam steering configuration.

\subsection{Performance Metric}
The primary performance metric is the \textit{complementary cumulative distribution function (CCDF)} of the relative received signal strength (RSS - relative in dB), defined as the probability that the RSS exceeds a predefined threshold \( \gamma_{\text{th}} \), formally,  $\mathrm{CCDF} = \mathbb{P}(\mathrm{RSS} \geq \gamma_{\text{th}}).$

The objective of the system is to maximize this probability for higher threshold levels by maintaining strong reflected mmWave links in NLoS conditions.

\begin{figure}[t]
    \centering
    \includegraphics[width=0.45\textwidth]{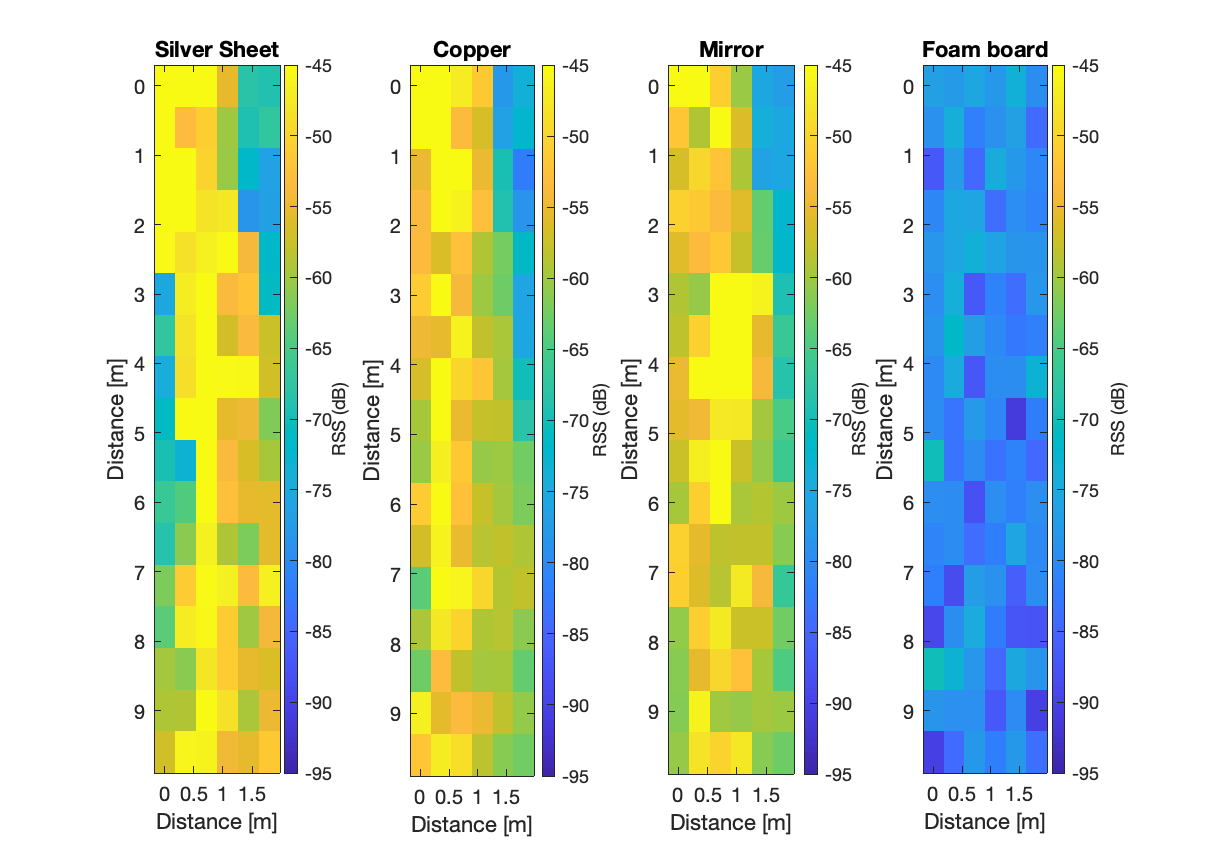}
    \caption{Total RSS values on 102 grid points obtained when using a silver
reflector, a copper reflector, a silver-coated mirror, and a foam board (no
reflector attached). } 
    \label{fig:heatmap2}
\end{figure}

\begin{figure}[t]
    \centering
    \includegraphics[width=0.45\textwidth]{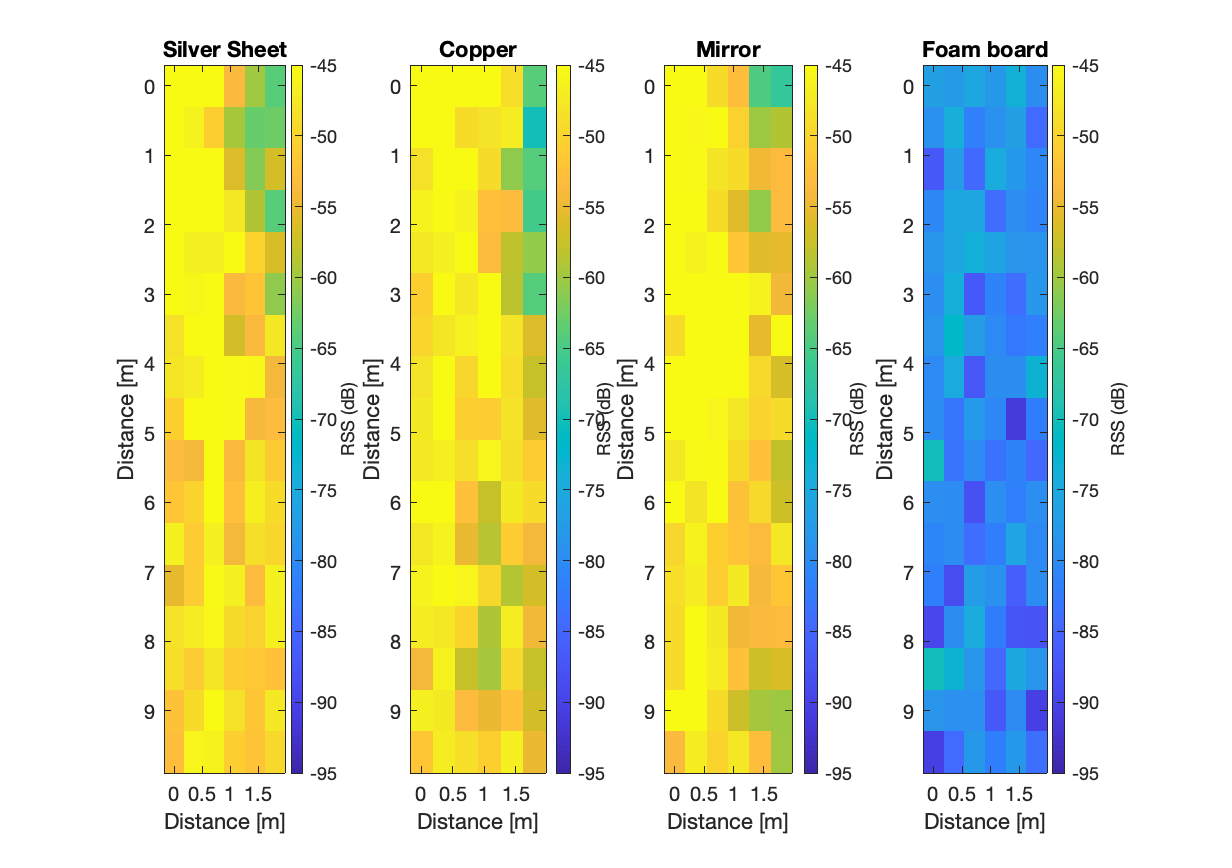}
    \caption{Total RSS measured  using adaptive beamforming with exhaustive angle search via passive reflectors.}
    \label{fig:heatmap3}
\end{figure}

\begin{figure}[t]
    \centering
    \includegraphics[width=0.85\linewidth]{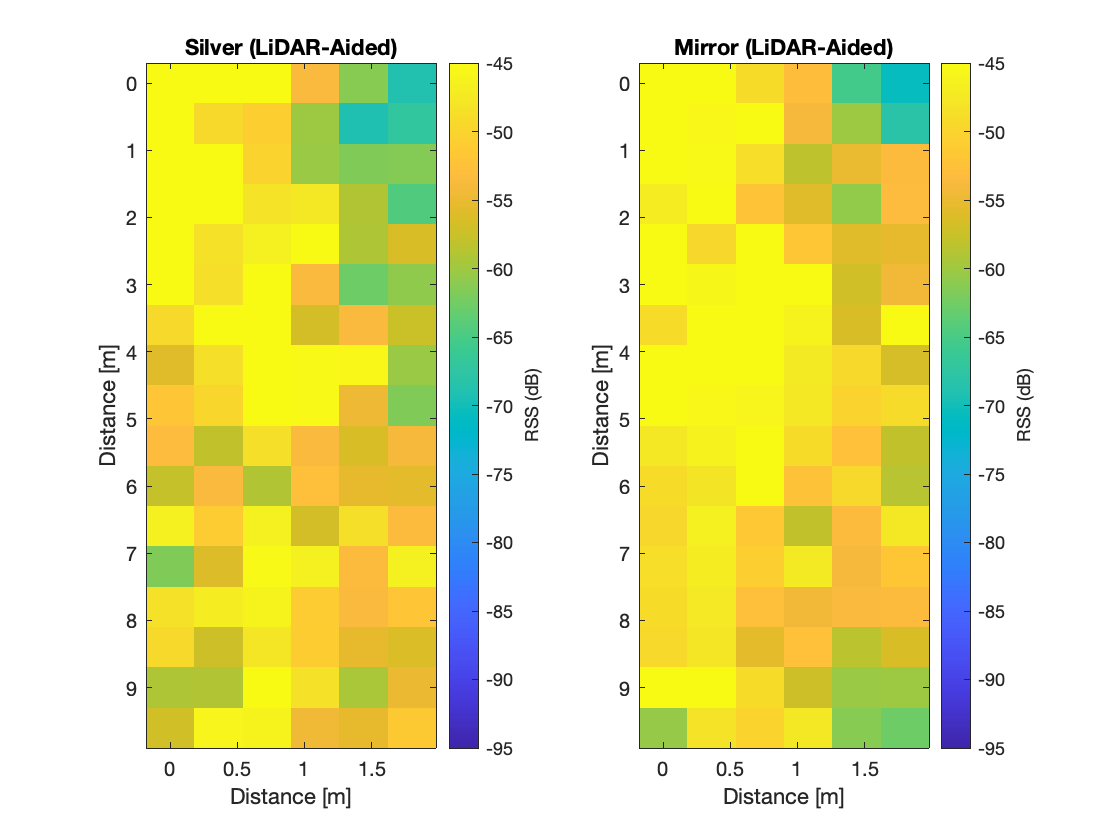}
    \caption{RSS measured at 102 grid points using LiDAR-driven beamforming in the NLoS region.  LiDAR derived AoDs were mapped to the nearest quantized mmWave AoDs. }
    \label{fig:hm}
\end{figure}

\begin{figure}[t]
\hspace{-7mm}
    \includegraphics[width=1.05\linewidth]{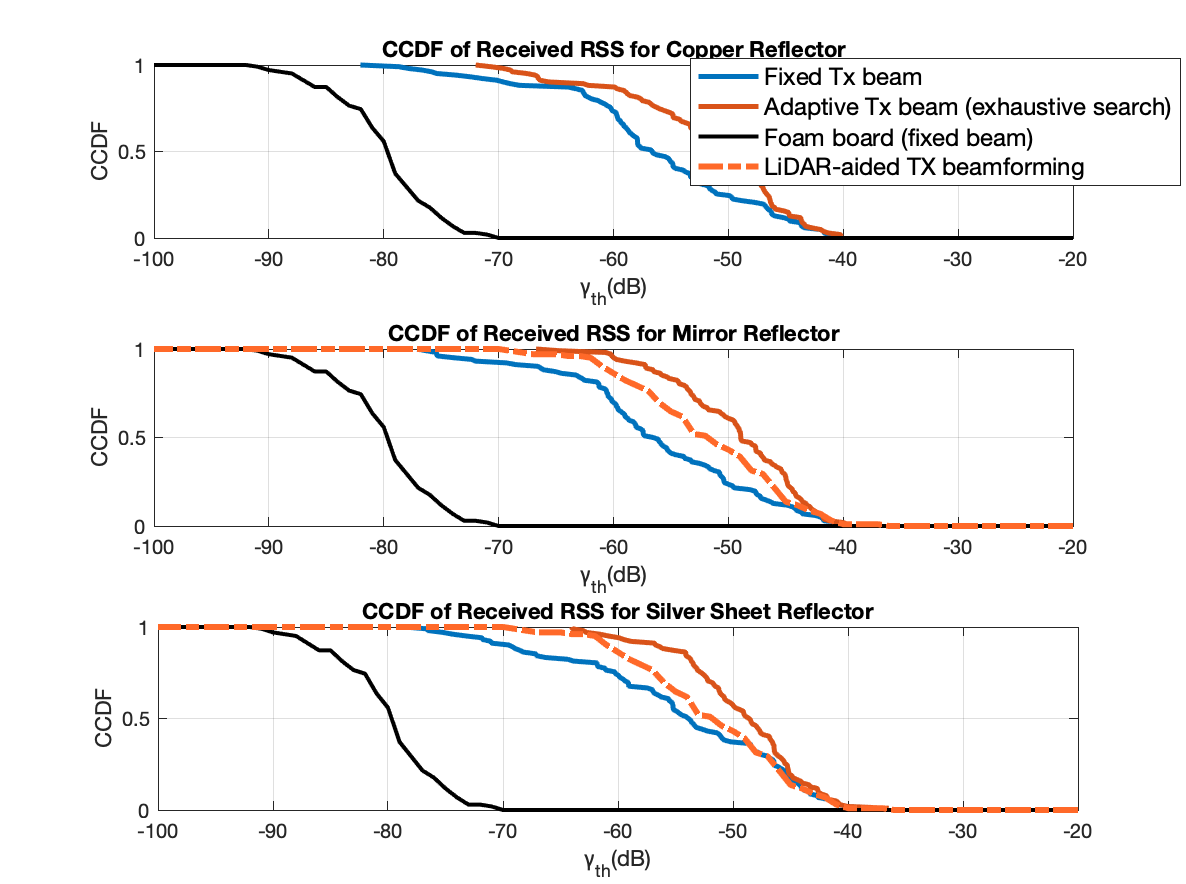}
    \caption{CCDF ($\text{Pr}(\text{RSS} > \gamma_{\text{th}})$) comparison of fixed vs. adaptive beamforming across reflector types.}
    \label{fig:ccdf}
\end{figure}


\section{Experimental Results}

This section evaluates the effectiveness of LiDAR-assisted adaptive beamforming and passive reflector integration in improving mmWave coverage under non-line-of-sight  conditions in indoor environments.

\subsection{Fixed Beam Performance}

Fig.~\ref{fig:heatmap2} shows the received signal strength distribution across the measurement grid using a fixed transmit beam with an angle of departure  set to $-5^\circ$ for each reflector. All tested reflectors,  silver sheet, copper, and mirror,  demonstrate significant improvements in signal strength compared to the foam board baseline.   However, fixed beam steering results in non-uniform coverage and notable signal degradation at several grid point locations.

\subsection{Adaptive Beamforming with LiDAR Assistance}
To address coverage gaps and enhance signal uniformity, adaptive beamforming was implemented. The transmitter scanned across multiple azimuthal angles of departure (AoDs),  \(0^\circ, \pm1.5^\circ, \pm3^\circ, \pm5^\circ, \pm10^\circ, \pm15^\circ\),  and selected the optimal beam direction for each grid point using two strategies: (i) exhaustive search and (ii) LiDAR-assisted visibility mapping. The resulting RSS distributions are shown in Figs.~\ref{fig:heatmap3} and~\ref{fig:hm}. Both methods significantly improved signal coverage, with LiDAR guidance achieving near-optimal beam selection without requiring exhaustive probing.  

The mirror reflector demonstrated superior performance due to improved LiDAR detection capabilities. In contrast, the glossy silver reflector exhibited limited LiDAR detection beyond 5~m, leading to intermittent (``spotty'') user visibility. Consequently, the LiDAR-assisted method defaulted to an AoD of \(0^\circ\) in these cases, bypassing optimization. This limitation stems from the semi-specular nature of the glossy silver surface, which reduces specular reflections necessary for reliable LiDAR sensing.  As shown in Fig.~\ref{fig:hm}, the mirror reflector enabled broader mmWave coverage thanks to more consistent user detection. Nonetheless, some coverage irregularities persisted, primarily due to multipath scattering at the receiver.

\subsection{CCDF-Based Performance Evaluation}

Fig.~\ref{fig:ccdf} presents the complementary cumulative distribution function of the RSS across all grid points. Compared to the fixed beam case, adaptive beamforming results in a rightward shift of the CCDF curves for all reflectors, indicating enhanced signal reliability. Specifically, at $\mathrm{Pr} = 1$, the RSS improves by approximately 12~dB for the silver reflector, 9~dB for the mirror, and 6~dB for copper. These gains confirm that adaptive beam selection enhances NLoS coverage.  Notably, LiDAR-guided beamforming achieves performance comparable to that of exhaustive search, validating its practical utility. However, a small performance gap remains which can be attributed to receiver-side multipath effects not fully captured by LiDAR. This suggests that while LiDAR enables effective initial beam selection, additional mechanisms such as environment fingerprinting may be required.

\subsection{LiDAR Detection Range}

Table~\ref{Ltable} summarizes the LiDAR detection performance for the examined reflectors.  Detection coverage is defined as the proportion of grid points (out of 102) where a user was successfully detected in the LiDAR point cloud. The mirror reflector yields the highest coverage at 95\%, followed by the glossy silver sheet at 70\%. Copper achieves only 4\% coverage, despite its strong mmWave reflectivity. This is due to its rough surface, which causes scattering and weakens LiDAR return.

\begin{table}[t]
\centering
\small
\caption{LiDAR Detection Coverage Across Reflector Types}
\label{Ltable}
\begin{tabular}{|c|c|}
\hline
\textbf{Reflector Type} & \textbf{LiDAR Detection Rate} \\
\hline
Mirror         & 95\% \\
Glossy Silver  & 70\% \\
Copper         & 4\%  \\
\hline
\end{tabular}
\end{table}

\section{Conclusion}
This paper presented a LiDAR-assisted beamforming framework that integrates large specular/semi-specular passive reflectors to enhance mmWave signal coverage in indoor NLoS environments. By combining environment-aware spatial sensing with adaptive beam steering, the system identifies optimal angles of departure without requiring receiver-side feedback. Experimental results across 102 grid locations demonstrate that co-reflective surfaces, when guided by LiDAR input, significantly improve link reliability under NLoS conditions. The proposed approach offers a scalable, energy-efficient, and low-cost solution for extending coverage in future 6G and beyond networks.  Future work will examine additional surface types and reflection point selection for multi-user NLoS communication.

\section*{Acknowledgment}
This material is based on work supported by the National Science Foundation under grant No. NSF-2243089.


\end{document}